# Dimensionality of mobile electrons at X-ray-irradiated $LaAlO_3/SrTiO_3$ interfaces

V. N. Strocov,[1] F. Lechermann,[2] A. Chikina,[1] F. Alarab,[1] L. L. Lev,[1,3] V. A. Rogalev,[1,4] T. Schmitt,[1] and M.-A. Husanu[1,5]

[1]Swiss Light Source, Paul Scherrer Institute, CH-5232 Villigen-PSI, Switzerland

[2]Institut für Theoretische Physik, Universität Hamburg, Jungiusstr. 9, DE-20355 Hamburg, Germany

[3]Moscow Institute of Physics and Technology, 9 Institutskiy lane, RU-141700 Dolgoprudny, Russia

[4]Julius-Maximilians-Universität, Physikalisches Institut, Am Hubland, 97074 Würzburg, Germany

[5]National Institute of Materials Physics, Atomistilor 405A, RO-077125 Magurele, Romania

Electronic structure of $LaAlO_3/SrTiO_3$ (LAO/STO) samples, grown at low oxygen pressure and post-annealed ex-situ, was investigated by soft-X-ray ARPES focussing on the Fermi momentum ($\mathbf{k}_F$) of the mobile electron system (MES). X-ray irradiation of these samples at temperatures below 100K creates oxygen vacancies ($V_O$s) injecting Ti $t_{2g}$-electrons into the MES. At this temperature the oxygen out-diffusion is suppressed, and the $V_O$s should appear mostly in the top STO layer. The X-ray generated MES demonstrates, however, a pronounced three-dimensional (3D) behavior as evidenced by variations of its experimental $\mathbf{k}_F$ over different Brillouin zones. Identical to bare STO, this behavior indicates an unexpectedly large extension of the X-ray generated MES into the STO depth. The intrinsic MES in the standard LAO/STO samples annealed in-situ, in contrast, demonstrates purely two-dimensional (2D) behaviour. Based on self-interaction-corrected DFT calculations of the MES induced by $V_O$s at the interface and in STO bulk, we discuss possible scenarios of the puzzling 3D-ity. It may involve either a dense ladder of quantum-well states formed in a long-range interfacial potential or, more likely, X-ray-induced bulk metallicity in STO accessed in the ARPES experiment through a short-range interfacial barrier. The mechanism of this metallicity may involve remnant $V_O$s and photoconductivity-induced metallic states in the STO bulk, and even more exotic mechanisms such as X-ray induced formation of Frenkel pairs.

## Introduction

Transition-metal oxides (TMOs) presently play one of the forefront roles in theoretical and experimental condensed matter research (for entries see [1]). An involved interplay between the spin, charge, orbital and lattice degrees of freedom in these materials results in a wealth of phenomena interesting from the fundamental point of view and bearing potential for technological applications. These include rich electronic and magnetic phase diagrams, metal-insulator transitions, colossal magnetoresistance, ferroelectricity, multiferroicity, high-$T_c$ superconductivity, etc. Interfaces and heterostructures of TMOs can add another dimension to their fascinating properties, giving rise to physical phenomena which cannot be anticipated from the properties of individual constituents, with the new functionalities having great promise for future device applications (see, for example, the reviews [2,3]).

The interface between $LaAlO_3$ (LAO) and $SrTiO_3$ (STO) is a paradigm example of new functionalities that can be formed by interfacing TMOs. Although bulk LAO and STO are both band insulators, their interface spontaneously forms a mobile electron system (MES) [2,3]. Its high electron mobility co-exists with superconductivity, ferromagnetism, large magnetoresistance and other non-trivial phenomena which can in addition be tuned with field effect. The MES electrons are localized in the interfacial quantum well (QW) on the STO side and populate the Ti $t_{2g}$-derived in-plane $d_{xy}$-states and out-of-plane $d_{xz/yz}$-states [4–6]. The latter, furthermore, can be manipulated through artificial quantum confinement in thin STO layers [7]. Whereas in stoichiometric LAO/STO the MES is localized within a few layers from the interface and is therefore purely 2D, oxygen deficiency of STO can extend the MES to a much larger depth of more than 1000 Å, resulting in its essentially 3D character [8–10]. Two phenomena bearing key importance for virtually all physical properties of LAO/STO are (1) polaronic nature of the charge carriers, where strong electron-phonon coupling to the LO3 phonon mode reduces their low-temperature mobility by a factor of about 2.5, and coupling to soft phonon modes dramatically reduces mobility with temperature [11,12], and (2) electronic phase separation (EPS) where the conducting MES puddles are embedded in the insulating host phase [12–14].

X-ray irradiation can dramatically change electronic and magnetic properties of oxide materials, which is typically connected with creation of oxygen vacancies ($V_O$s) [14–22]. In a simplified picture of oxygen-deficient (OD) STO, each $V_O$s releases two electrons from the neighbouring Ti atom. One of them joins the MES formed by delocalized quasiparticles, which are Ti $t_{2g}$ derived, weakly correlated, non-magnetic and form large polarons [11]. The other electron stays near the Ti ion to form localized in-gap state (IGSs) at binding energy $E_B$ ~ -1.3 eV, which are Ti $e_g$ derived, strongly correlated, magnetic and are often viewed as small polarons. In a more elaborate picture, the electron distribution between the MES and IGSs depends on particular configurations of $V_O$s [23] which tend to cluster [14]. The $V_O$s have a dramatic effect on the transport properties of STO, with their concentration of only 0.03% transforming STO into metal [24]. This picture of dichotomic electron system in the bulk as well at the surfaces and interfaces of OD-STO can be described within the combination of density functional theory with explicit electron-correlation schemes, such as e.g. dynamical mean-field theory (DMFT) (see, for example, Refs. [21,25–29]). Recently, it has been experimentally confirmed by resonant photoemission (ResPE) [22]. The coexistence of the two radically different MES and IGS electron subsystems hugely enriches the physics of the OD-STO systems compared to the stoichiometric ones, critically affecting the conductivity, magnetism and EPS.

Here, we use soft-X-ray angle-resolved photoelectron spectroscopy (ARPES) to explore the dimensionality of the MES generated by X-ray irradiation of LAO/STO heterostructures grown in oxygen-deficient conditions and post-annealed (PA-LAO/STO). Our analysis focuses on variations of the experimental $\mathbf{k}_F$ across the Brillouin zones (BZs), with the $\mathbf{k}_F$ values determined from the gradient of bandwidth-integrated spectral intensity. Although the $V_O$s generated by X-rays at our low temperature of 12K should be located in the top STO layer, we observe an unexpected 3D-ity of X-ray generated MES, indicating its large extension into the STO depth. This behaviour, identical to the X-ray irradiated bare STO, contrasts with the 2D behavior of the intrinsic MES in standard LAO/STO samples. Based on self-interaction-corrected DFT calculations, we discuss possible scenarios how the 3D-MES is formed.

## Sample preparation

Our LAO/STO samples with a LAO layer of the critical thickness 4 u.c. on top of $TiO_2$-terminated STO(100) were grown with Pulsed Laser Deposition. We followed a non-standard growth protocol, where the STO substrate was annealed at 500°C in vacuum, the $O_2$ pressure during the LAO deposition at 720°C reduced to $8\times10^{-5}$ mbar, and the post-growth annealing for 12 hrs performed ex-situ at a temperature of 500°C [14,22]. Although such samples are finally (nearly) stoichiometric in terms of the chemical-element ratio, as evidenced by their photoelectron spectroscopy characterization discussed below, their transport properties [30] are quite different from the LAO/STO samples grown and annealed under the standard protocol [11]. Their difference most important in the context of our ARPES study is that under X-ray irradiation at temperatures below ~100K the post-annealed samples readily built up $V_O$s in STO. In the ARPES spectra acquired in parallel with the irradiation, this fact is evidenced by gradual development of the initially absent IGS peak in the band gap and $Ti^{3+}$ component of the Ti 2*p* core levels, which are characteristic signatures of the $V_O$s (for detailed analysis of the time evolution see [14]).

The exact mechanism of irradiation-induced creation of $V_O$s at the LAO/STO interfaces and even bare STO surfaces is not yet quite understood. There are a few intriguing observations: (1) As unfolded below in this paper, the X-ray generated MES is 3D, suggesting its expansion over a significant depth into STO; (2) The X-ray generated $V_O$s and corresponding MES can be observed only at low sample temperatures below ~100K. Once created, they stay stable without further irradiation for at least tens of hours. Upon increase of temperature well above ~100K, however, they quench and can not develop even at relatively high synchrotron-radiation power densities (see below). This fact excludes that the classical thermally-activated diffusion of oxygen atoms out of STO be involved in the generation of the $V_O$s, because in this case the temperature dependence would be opposite. Instead, their creation and/or stabilization may in some way be linked to the cubic to tetragonal phase transition and concomitant creation of domains in STO below 105 K [31,32]. The electric field in STO due to the interfacial band bending, separating the photoexcited electrons and holes [33] as well as the repulsion of the mobile photoexcited electrons from the localized ones staying at the $V_O$s [34] may also contribute to the stability of the X-ray generated $V_O$s and MES; (3) The $V_O$s can be quenched by exposure of the LAO/STO samples to X-ray irradiation in $O_2$ pressure of about $10^{-7}$ mbar on a time scale of seconds, which is much faster than the generation of $V_O$s on a time scale of tens of minutes. The high efficiency of this reaction can be explained by that X-ray irradiation cracks the physisorbed $O_2$ molecules into atomic oxygen [19,20]. We note that whereas for STO surfaces only absorption of $O_2$ molecules is sufficient to quench the $V_O$s [17], in our case of the buried LAO/STO interface their X-ray cracking is necessary. This observation can be explained, tentatively, by that oxygen would effectively penetrate through the LAO layer only in its atomic form; (4) The $V_O$-generation rate strongly depends on the sample growth protocol, including temperature and $O_2$ pressure during the substrate annealing and the LAO deposition, and in-situ vs ex-situ annealing. The LAO/STO samples grown under the standard protocol (ST-LAO/STO), including substrate annealing in $O_2$, high $O_2$ pressure during the LAO deposition and in-situ annealing, are fairly immune to X-ray irradiation [11]. The susceptibility of the post-annealed samples to X-ray irradiation may be connected with certain amount of structural defects remaining in them after the post-annealing despite their complete or nearly complete stoichiometric recovery. At the moment, it is difficult to reconcile all these observations in one single mechanism of creation and stabilization of the $V_O$s, and this is not a prime objective of this study.

## Experimental and theoretical methods

Our SX-ARPES experiments used ResPE at the Ti *L*-edge in order to boost the signal from the Ti derived electron states at the buried LAO/STO interface. The measurements were performed at the soft-X-ray ARPES endstation [35] of the Advanced Resonant Spectroscopies (ADRESS) beamline [36] at the Swiss Light Source, Paul Scherrer Institute, Switzerland. A photon flux of about $10^{13}$ photons/sec was focused in a spot size of 30 x 75 $\mu m^2$ on the sample surface, and the combined (beamline and analyzer) energy resolution ($\Delta E$) was ~50 meV full width at half maximum (FWHM). The sample was kept at 12 K in order to suppress smearing of the **k**-dispersive spectral fraction [37] and, most importantly, allow a build-up of the $V_O$s under X-ray irradiation (see above). The X-ray absorption spectroscopy (XAS) data were measured in the total electron yield (TEY). All ARPES and XAS data presented below are acquired at saturation after more than 2 hrs of irradiation time. The XAS and angle-integrated photoemission data were measured with circularly polarized incident X-rays. The in-plane Fermi surface (FS) maps and ARPES band dispersion images presented below were measured with *s*-polarized X-rays, and the out-of-plane FS map with *p*-polarized X-rays.

Our electronic structure calculations for oxygen-deficient (OD) LAO/STO interfaces addressed possible MES contributions from $V_O$s in the top $TiO_2$ layer as well as in deeper-lying $TiO_2$ layers on equal footing as embedded in a LAO/STO supercell. The calculations utilized the STO-bulk lattice constant of a=3.905 Å, and were structurally relaxed within the generalized-gradient approximation (GGA) of Kohn-Sham's Density-Functional Theory (DFT) utilizing a mixed-basis pseudopotential code. The final electronic structure was determined with a DFT+self-interaction-correction (SIC) scheme, where explicit Coulomb interactions on oxygen sites are described within SIC, and are incorporated in the O pseudopotential [38]. The SIC was applied to the O(2*s*) and the O(2*p*) orbitals via weight factors $w_p$ (see Ref. [38] for more details). While the O(2*s*) orbital is by default fully corrected with $w_p$=1.0, the common choice $w_p$=0.8 was used for O(2*p*) orbitals. Since the Ti $t_{2g}$ states, mostly relevant for the MES, are only weakly occupied in LAO/STO, neglecting the explicit Coulomb interactions on Ti when focussing on the MES proves as an adequate approximation for the very large supercells. Note that when applied to bulk STO, the present DFT+SIC treatment results in a reasonable band gap of 3 eV, curing the notorious underestimation in standard exchange-correlation functionals.

## Electronic structure overview

Fig. 1 (*a*) shows the Ti $2p_{3/2}$ core-level spectrum of our PA-LAO/STO samples taken at $h\nu$ = 1000 eV, which is decomposed into the $Ti^{4+}$ and $Ti^{3+}$ components (linear background removed). The latter is characteristic of the $V_O$s. (*b*) shows XAS spectrum through the Ti $L_3$- and $L_2$-edges (excited from the $2p_{3/2}$ and $2p_{1/2}$ core levels, respectively). The pairs of salient XAS peaks at each edge are formed by the $Ti^{4+}$ $e_g$ and $t_{2g}$ states. The corresponding map of ResPE intensity as a function of $E_B$ and excitation energy $h\nu$, identifying the Ti-derived electron states [39–43] is displayed in Fig. 1 (*c*). Above the broad valence band composed from the O 2*p* states in STO and LAO, the map shows the resonating broad IGS peak at $E_B$ ~ -1.3 eV which is a hallmark of the $V_O$s. Recent analysis of its resonant behavior [22] confirms that the IGS-subsystem is derived from $e_g$ states of the $Ti^{3+}$ ions pushed down in energy by strong electron correlations [25,27,28]. The narrow resonating peak at the Fermi energy ($E_F$) identifies the $t_{2g}$-derived MES. Its intensity maximum is shifted from the $Ti^{4+}$ $t_{2g}$ to $Ti^{4+}$ $e_g$ peak in the XAS spectrum due to remnant **k**-conservation between the intermediate and final states coupled in the ResPE process [22].

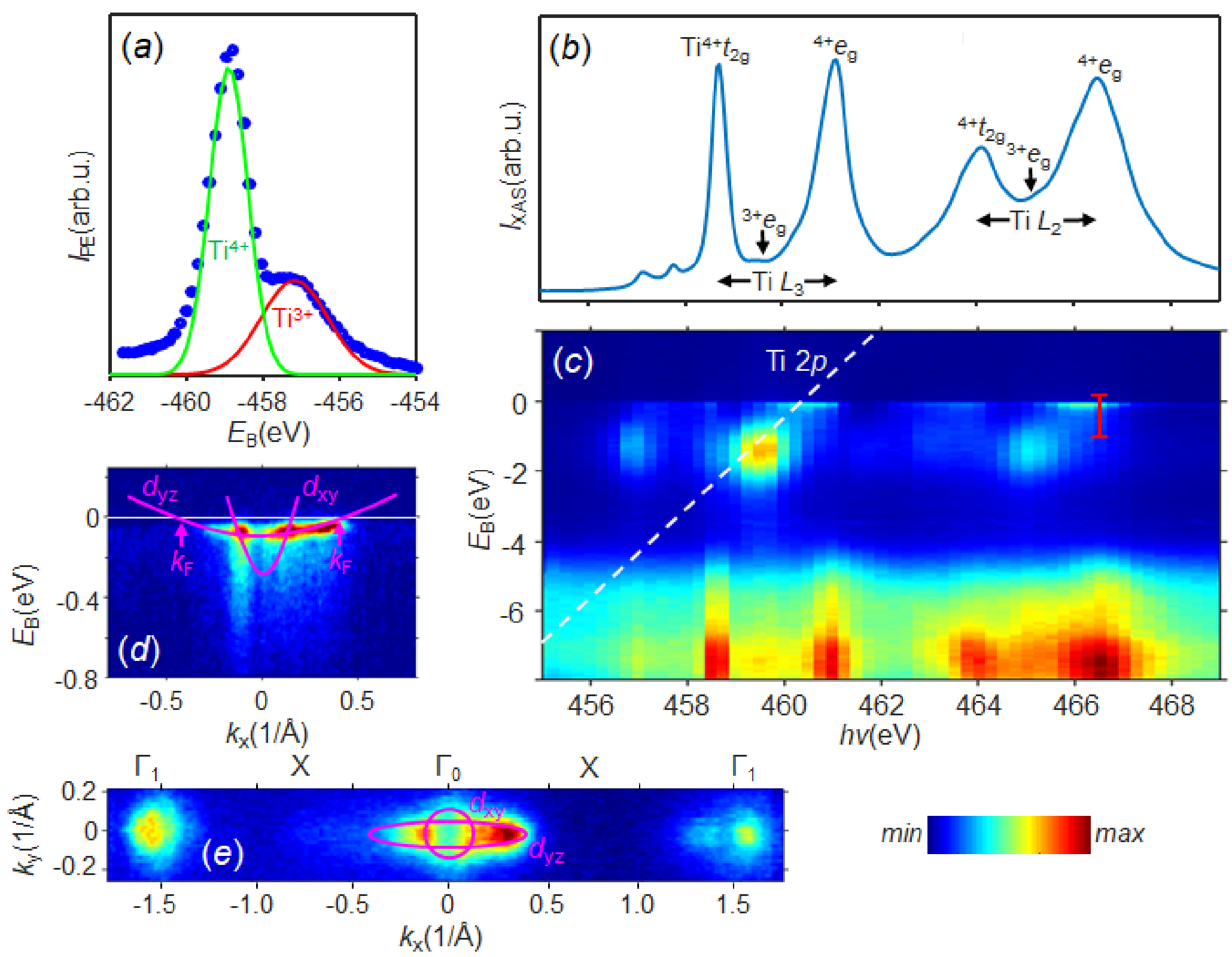

**Fig. 1**. Spectroscopic overview of PA-LAO/STO: (*a*) Ti $2p_{3/2}$ core level decomposed into the $Ti^{4+}$ and $Ti^{3+}$ components; (*b*) XAS spectrum and (*c*) angle-integrated ResPE intensity through the $L_3$ and $L_2$ edges. The resonating peak at $E_F$ signals the $t_{2g}$-derived MES and the one at $E_B$~ -1.3 eV the IGS characteristic of the $V_O$s; (*d*) the ARPES image along the ΓX direction measured at $h\nu$ = 466.4 eV as marked by red bar at (*c*); (*e*) FS measured at the same $h\nu$. The $d_{xy}$ and $d_{yz}$ states, selected by the *s*-polarized X-rays, are indicated.

We note that $Ti^{3+}$ $e_g$ and $t_{2g}$ signals in the XAS spectrum in Fig. 1 (*b*), corresponding to the Ti atoms hosting the $V_O$s, are almost invisible. At the same time, the IGS signal in the ResPE map (*c*), corresponding to the $Ti^{3+}$ $e_g$ states of these atoms, is profound. This observation shows that most $V_O$s are located in vicinity of the top STO layer, because the width of this region is much smaller than the probing depth of the TEY-XAS measurements of the order of 50 Å, and comparable with that of SX-ARPES measurements of the order of 11 Å at our excitation energy [20].

The image (*d*) in Fig. 1 shows the ARPES intensity measured along the ΓX direction with *s*-polarized incident X-rays, selecting predominantly the antisymmetric $d_{xy}$- and $d_{yz}$-derived electron states [6,11,22]. Furthermore, the choice of $h\nu$ = 466.4 eV at the $L_2$ edge enhanced the $d_{yz}$ states in comparison with the $d_{xy}$ ones [22]. The ARPES data is overlaid with a sketch of the $d_{xy}$ and $d_{yz}$ dispersions. We note pronounced spectral intensity variations along the bands, caused by energy- and **k**-dependent photoemission matrix elements. Particularly remarkable is the spectral intensity asymmetry relative to the Γ points, absent in VUV-ARPES on bare STO surfaces [18,44]. With *s*-polarized X-rays in our experimental geometry, this phenomenon is beyond the conventional dipole approximation for the photoemission process (because the angle between the photoemission direction and electric vector of the light is the same) and should be attributed to the photon momentum that is ~0.24 $Å^{-1}$ for our $h\nu$ in the soft-X-ray energy range. We observe that for PA-LAO/STO the lowest $d_{xy}$ band is much smeared compared to STO, and manifests itselves as intensity hot spots where it intersects the $d_{yz}$ band and hybridizes with it because of the symmetry breaking caused by the tetragonal lattice distortion and spin-orbit coupling. In view of the location of these $d_{xy}$ states in the top $TiO_2$ layer, the observed smearing should be connected with a disorder induced in this layer by the LAO overlayer and X-ray generated $V_O$s. As we will see below, the absence of high-order $d_{xy}$ bands in PA-LAO/STO is consistent with the steeper electrostatic band-bending potential $V(z)$ compared to bare STO caused by the additional electron density injected by the $V_O$s. We note that the band order and band dispersions observed in PA-LAO/STO (and STO) and ST-LAO/STO samples [6,11] demonstrate the same $d_{xy}<d_{xz/yz}$ band order. However, the observed band filling in PA-LAO/STO (STO) is somewhat larger than in ST-LAO/STO. Furthermore, the polaronic-hump weight and thus renormalization of the effective mass $m^*$ in PA-LAO/STO (STO) reduce to ~1.5 compared to ~2.5 in ST-LAO/STO [11]. Importantly, as we will see below, these materials demonstrate different dimensionality of the MES. Finally, Fig. 1 (*e*) shows the experimental FS of PA-LAO/STO where we observe the circular $d_{xy}$ and elliptical $d_{yz}$ sheets.

## Dimensionality of the MES

Do the MES electrons in PA-LAO/STO, developing under irradiation, stay confined at the interface and keep their 2D character, or expand into the STO bulk and become 3D? To answer this question, we investigated their ARPES response as a function of out-of-plane momentum $k_z$ varied in an extended $h\nu$-range above the Ti *L*-edge. Because of the much weaker off-resonance response of the MES, such measurements are extremely photon hungry. We used *p*-polarization of incident X-rays to switch from the two-band combination of the $d_{xy}$ and $d_{yz}$ antisymmetric states to the single-band $d_{xz}$ state. Fig. 2 shows an experimental map of ARPES intensity at $E_F$ as a function of $h\nu$, rendered into the out-of-plane momentum $k_z$ according to the relation $k_z = \sqrt{\frac{2m\left(E_k+V_0\right)}{\hbar^2} - k_{//}^2} + \frac{p_z}{\hbar}$, where $m$ is the free-electron mass, $E_k$ photoelectron kinetic energy, $V_0$ the inner potential set to 15 eV, and $p_z$ the photon momentum correction in our experimental geometry [35].

The map in Fig. 2 clearly shows FS contours of the $d_{xz}$-derived interfacial states whose intensity periodically blows up when $k_z$ hits the Γ-points of the BZ. However, this pattern alone does not necessarily indicate a 3D-ity of these states, because 2D states formed by the out-of-plane $d_{xz}$-orbitals confined in the interfacial QW would also produce periodic intensity oscillations. The only difference is that the out-of-plane dispersion of the 3D states will manifest itself as rounded FS contours in the ($k_x$,$k_z$) coordinates, and the absence of it for the 2D states as straight FS contours (for in-depth analysis of ARPES response of 2D states see [45]). In our case, relatively poor statistics and variations of the photoemission matrix element with *hv* and $k_x$ do not allow reliable discrimination between these 2D and 3D patterns of the $k_z$-dependent ARPES intensity from the MES. As the acquisition of this dataset has already required more than 5 hrs, further increase of the acquisition time for these off-resonance measurements is less practical.

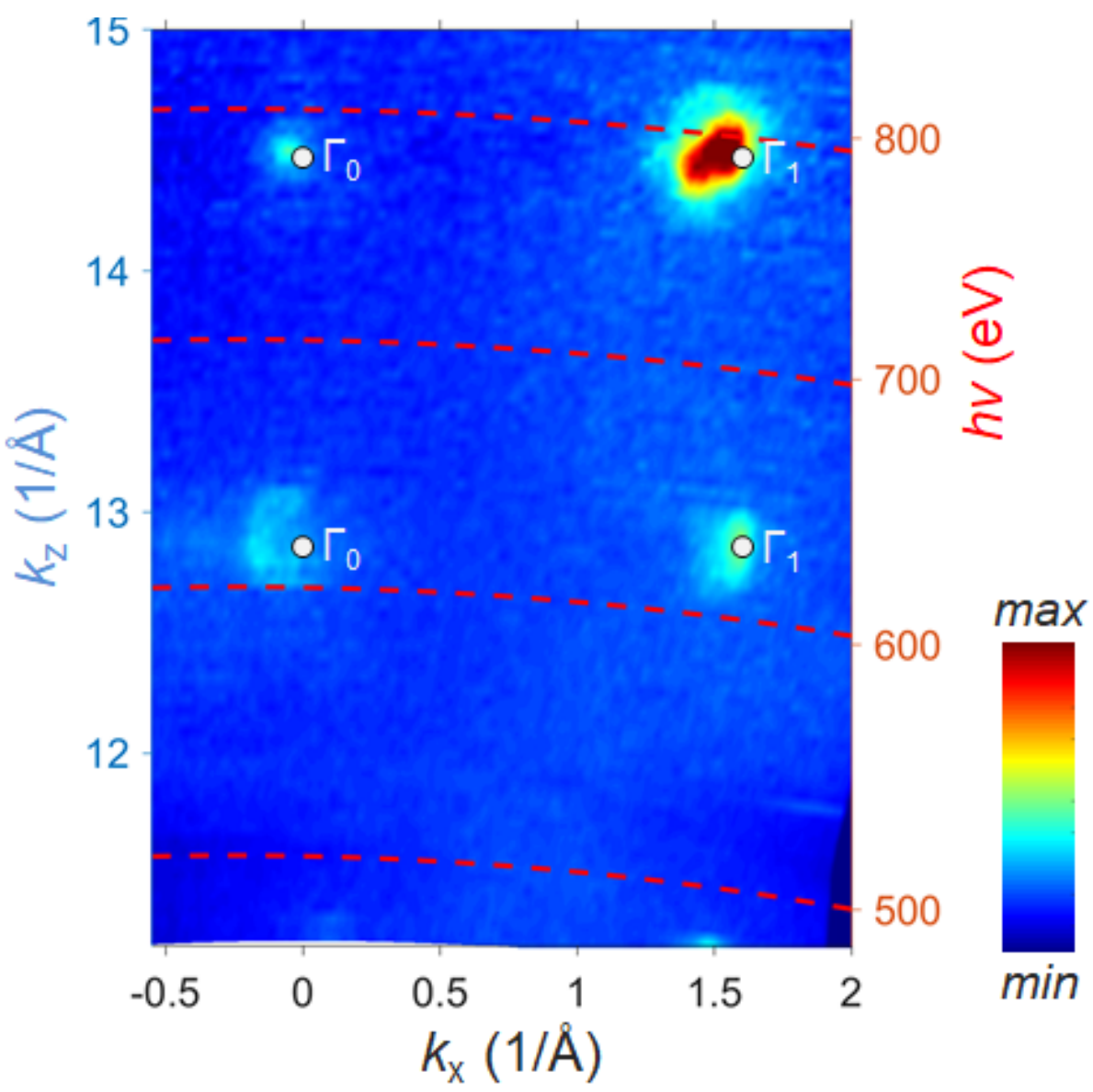

**Fig. 2.** The $d_{xz}$-derived FS in PA-LAO/STO as a function of $k_z$ in an extended *hv*-range. The corresponding *hv*-values are marked. The intensity is normalized to the angle-integrated intensity for each *hv*. The 3D-ity of the out-of-plane MES states, suggested by rounded FS contours, is confirmed by the $k_F$ variations across the BZs in Fig. 3.

In order to confirm the 3D-ity of the X-ray induced MES, we returned to the resonant *hv* = 466.4 eV delivering maximal intensity to the $d_{yz}$-states, and investigated the variation of the apparent $k_F$ as a function of $\mathbf{k}_{//}$ when moving to the next BZ. The measured ARPES image through two BZs is presented in Fig 4 (a). As in our case the occupied bandwidth $W$ ~ 50 meV is basically equal to the experimental $\Delta E$, accurate determination of $k_F$ is not trivial. Our ARPES simulations, described in the Appendix, have demonstrated that whereas the conventional method to determine $k_F$ from the peaks of spectral intensity at $E_F$ (intensity method) stays accurate only as long as $W$ is much larger than $\Delta E$, the extremes of the gradient $dI_W/dk$ of the bandwidth-integrated spectral intensity (gradient method) yield accurate $k_F$ values even when $W$ is of the order of or even smaller than $\Delta E$. The experimental $dI_W/dk$ (in our case integrated over an energy window of 100 meV, with the excluded polaronic introducing a negligible correction) is shown in Fig. 3 (b). Relevant for our MES dimensionality analysis is the $d_{yz}$-band, where the d$I_W$/d$k$ extremes marked $k_F^0$ and $k_F^1$ identify its apparent $k_F$ (for the $k_F$-determination method see Appendix). Remarkably, it reduces from $k_F^0$ ~ 0.44 Å$^{-1}$ in the first BZ to $k_F^1$ ~ 0.29 Å$^{-1}$ in the second one (hereinafter the accuracy of all $k_F$ values, including the statistical and instrumental errors, is about ±0.025 Å$^{-1}$). We note that on the left side of $\Gamma_0$ and right side of $\Gamma_1$ the spectral intensity from this band vanishes before reaching $E_F$, presumably due to a sharp variation of the photoemission matrix elements with **k**, and can not be used to determine $\mathbf{k}_F$. For another PA-LAO/STO sample, where the post-growth annealing time was increased to 24 hrs, we found a somewhat smaller $k_F^0$ ~ 0.40 Å$^{-1}$ that anyway reduced to $k_F^1$ ~ 0.29 Å$^{-1}$ in the second BZ. Furthermore, we have compared our results on PA-LAO/STO with previous resonant SX-ARPES measurements on bare STO (Supplementary in Ref. [22]) and have found in this reference system remarkably similar $k_F^0$ ~ 0.40 Å$^{-1}$ and $k_F^1$ ~ 0.29 Å$^{-1}$.

This difference of the apparent $k_F$ between the two BZs is actually a clear manifestation of the 3D character of the $d_{yz}$ band in both PA-LAO/STO and STO. This situation is sketched in Fig. 3 (c), showing the $d_{yz}$-derived FS sheet measured in our experiment. In this case the $k_{//}$-coordinate of $\mathbf{k}_F$ indeed depends on $k_z$, which decreases as $k_{//}$ increases into the next BZ. The effect of the relatively small change of $k_z$ between the two BZs on the apparent $k_F$ is amplified by the strong elongation of the FS sheets along the $k_x$-axis, with light- and heavy-electron axes being ~0.12 and 0.8 A$^{-1}$, respectively, as measured in the previous polarization-dependent studies on for PA-LAO/STO [14,22] and STO [17,18]. We note that our $h\nu$ = 466.4 eV at the Ti $L_2$-edge resonance corresponds to $k_z$= 6.99 (2π/$a$), i.e. brings $k_z$ almost ideally to the Γ-point of bulk STO where the MES is located. Had the $L$-edge intra-atomic resonance energy not coincided with the Γ-point, the spectral intensity from the MES would have hugely dropped, and the ARPES data acquisition would have required much longer acquisition time similarly to the dataset presented in Fig. 2.

Our SX-ARPES results on PA-LAO/STO and STO can be compared to the VUV-ARPES results of Plumb *et al*. [44] on STO surfaces prepared by annealing in vacuum. In this case, the emergent MES can be related to the $V_O$s only [14–22], without any intrinsic polar-discontinuity contribution. In line with our findings, the experimental $k_z$-dispersions measured in an $h\nu$ range below 100 eV have also demonstrated the 2D-ity of the Ti $d_{xy}$ states and 3D-ity of the $d_{xz/yz}$ ones, with the latter forming closed contours in the out-of-plane FS map [44]. An indirect confirmation of the 3D character of the X-ray generated MES is an observation that such MES has a higher mobility compared to the intrinsic one [33] that can be attributed to its extension from the interface into the defect-free STO bulk.

Remarkably, the ST-samples show a qualitatively different pattern of the ARPES variations with $k_{//}$. Fig. 4 (*a*) represents an ARPES image from the dataset of Ref. [11]. This dataset was acquired at the same experimental conditions as above. Although the spectral intensity of the $d_{yz}$-bands again strongly varies through the BZs, the corresponding d$I_W$/d$k$ extremes (*b*) show constant $k_F$ ~ 0.33 Å$^{-1}$, identifying the 2D-ity of the MES. In passing, we note a significantly larger polaronic renormalization of the $d_{yz}$-dispersion at the ST-LAO/STO interfaces compared to PA-LAO/STO, where the electron-phonon interaction is screened by the larger MES density. Also the two hot spots of the ARPES intensity, where the $d_{xy}$ band intersects the $d_{yz}$ one, are more pronounced for ST-LAO/STO; the larger smearing of the $d_{xy}$ states in PA-LAO/STO can manifest a larger disorder in its top STO layer due to the random X-ray generated $V_O$s in this region.

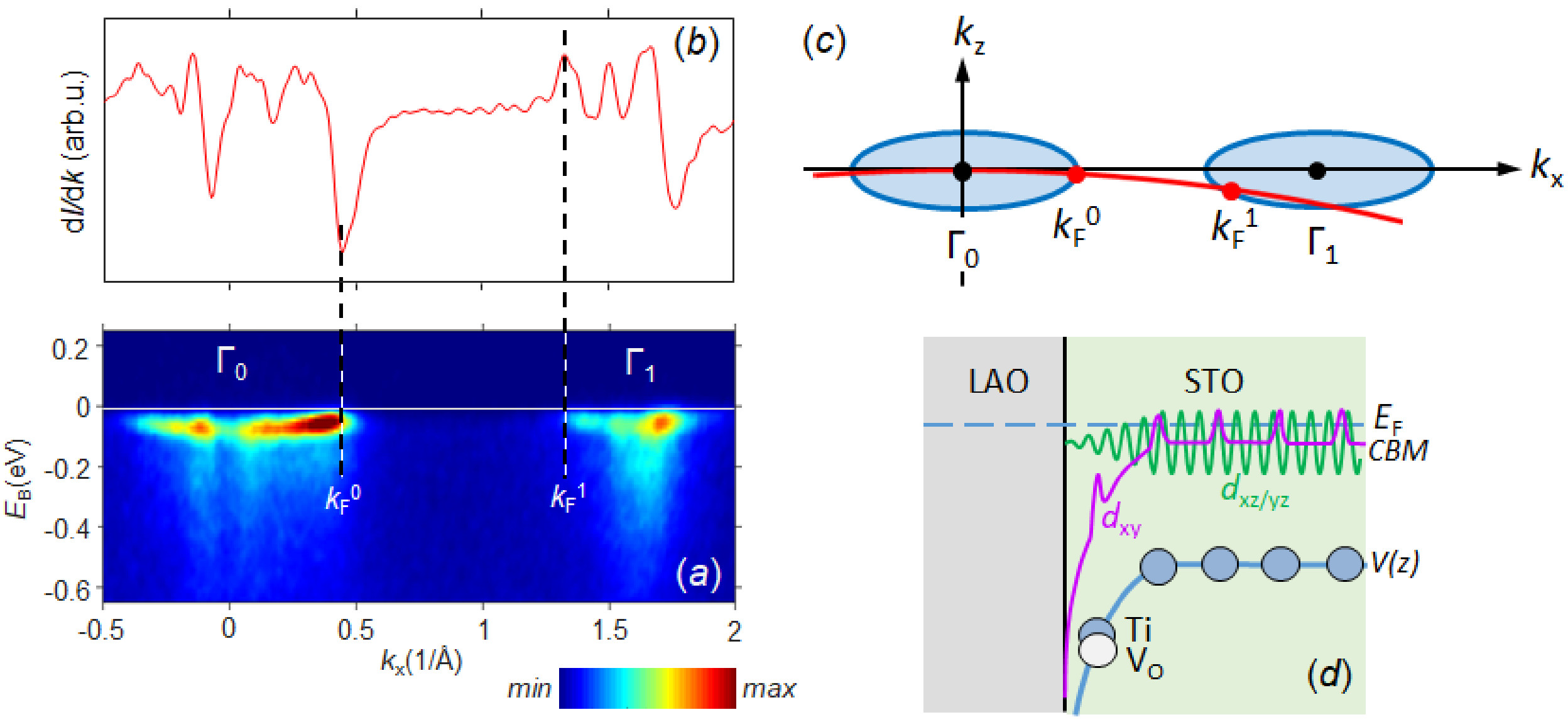

**Fig. 3.** 3D-MES in PA-LAO/STO: (*a*) Resonant ARPES image through two BZs measured at $h\nu$ = 466.4 eV; (*b*) corresponding gradient d$I_W$/d$k$ of the bandwidth-integrated intensity, where the dashed lines mark $k_F$ of the $d_{yz}$ bands. As sketched in (*c*), the difference in the apparent $k_F$ of the $d_{yz}$ bands between the two BZs identifies the out-of-plane dispersion and thus 3D-MES; (*d*) Schematics of the sharp interfacial band bending, screened by the additional electrons supplied by the $V_O$s, and of the corresponding wavefunctions of the $d_{xy}$ and $d_{xz}$ states. The $t_{2g}$ electrons, propagating into the STO bulk, form the 3D-MES.

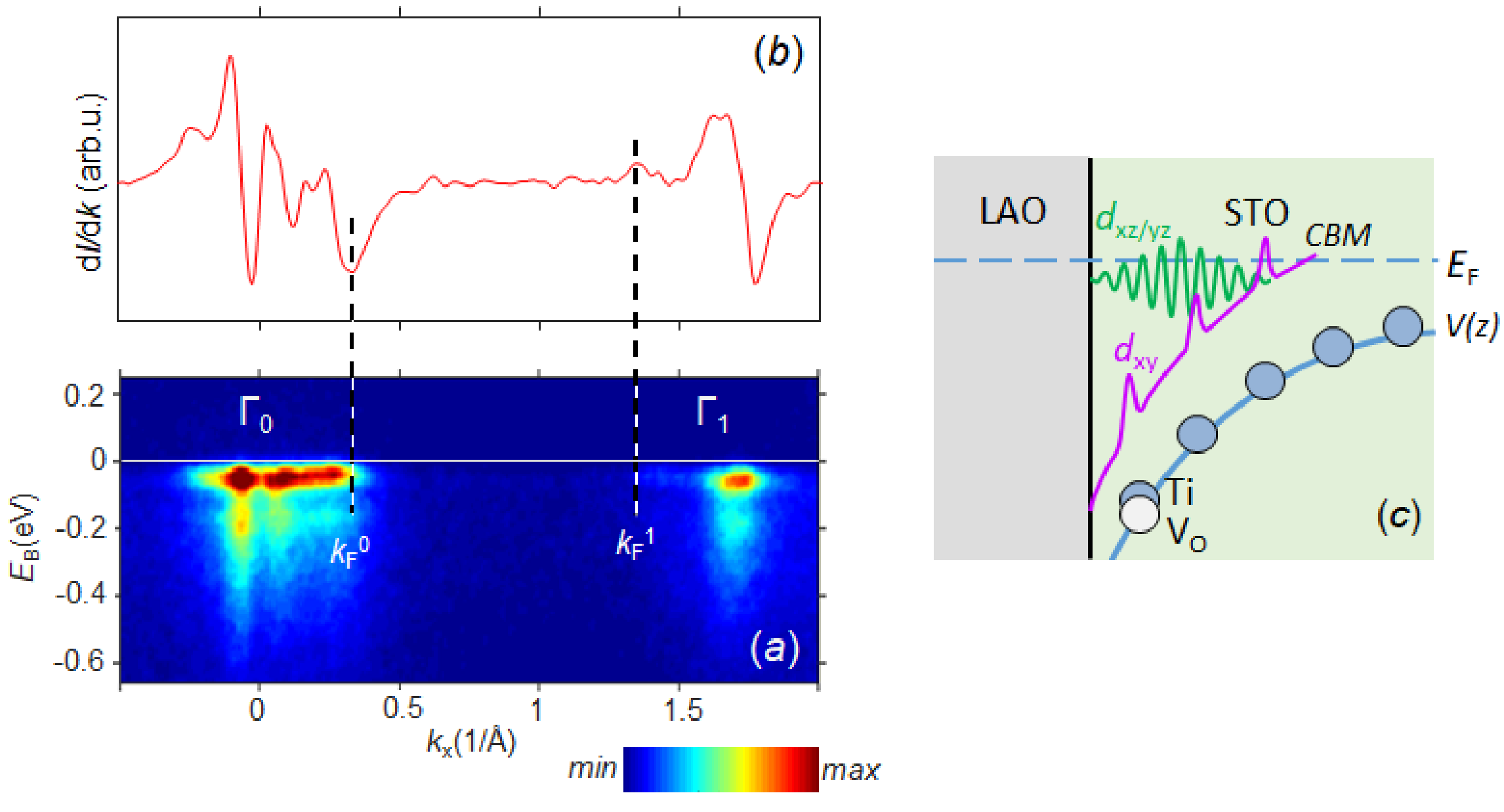

**Fig. 4.** 2D-MES in ST-LAO/STO: (*a*) Resonant ARPES image through two BZs measured at $h\nu$ = 466.4 eV; (*b*) d$I_W$/d$k$ plot, where the dashed lines mark $k_F$ of the $d_{yz}$ bands. Its identical value between the two BZs identify pure 2D-ity of the MES in ST-LAO/STO; (c) Schematics of an interfacial band bending embedding the $d_{xy}$ and quantum-confined $d_{xz/yz}$ states which form the 2D-MES.

We note that recent results by Soltani *et al*. [46] suggest that a 2D-MES may also be realized at bare STO surfaces. This has been identified from a splitting of the $d_{xz/yz}$-derived states, suggesting their quantum confinement and thus 2D-ity. Obviously, the dimensionality of the MES is determined by the depth and extension of the surface/interfacial QW which, in turn, is sensitive to the stoichiometry and preparation of the samples, which was in-situ cleavage in the experiments by Soltani *et al*. [46] vs in-situ annealing in the experiments by Plumb *et al*. [44]

## Discussion

### *Bulk and interfacial oxygen deficiency in STO*

The $V_O$s, developing under X-ray irradiation, should be one of the key factors of the observed 3D-ity of the MES in PA-LAO/STO, in contrast to the conventional 2D-MES in ST-LAO/STO. First of all, we note that LAO/STO interfaces grown on top of OD-STO have previously been studied by a number of techniques other than ARPES such as magnetotransport [8] and conducting-tip atomic force microscopy of the interfacial cross-section [9]. It has been noted that their $n_s$ can be two-three orders of magnitude larger than $n_s$ = 0.5 e per u.c. area predicted by the electrostatic arguments. This fact alone necessitates the existence of an extended and thus 3D component of the MES in OD-samples, and it has indeed been observed that sufficiently high concentrations of $V_O$s result in a dimensionality transformation of the MES from 2D to 3D [8–10]. Electron mobility in these systems could significantly exceed that of the paradigm 2DES in ST-LAO/STO [8,30,47]; this phenomenon could trace back to stronger electron screening of the polaronic interactions in LAO/STO by larger electron density [11]. However, the $V_O$ distribution profile in these OD-samples would extend into STO by a few thousands of Å and more [48], also affected by oxygen trapping and diffusion at domain walls (see, for example, Refs. [49–51]).

In contrast to these initially-OD samples, the post-annealing of our samples in oxygen, albeit ex-situ, ensures that before the X-ray irradiation they were at most stoichiometric. As discussed above, this fact is evidenced by the initial absence of the IGS peak and $Ti^{3+}$ core-level component in the ARPES spectra. The X-ray generated $V_O$s should then locate mostly in the top layer of STO because, while easy out-diffusion of oxygen atoms from STO through the LAO overlayer can be explained by typically relaxed crystallinity of the latter [52], the out-diffusion from deeper layers of STO should be practically prohibited due to vanishing diffusion coefficient of oxygen at our low sample temperature. Our case should therefore be different from creation of $V_O$s by deposition of thin metal films on STO at higher temperature where the $V_O$s could be distributed over a depth of ~1 nm [53,54]. This top-layer location of the $V_O$s in our case is consistent with the above observation that the $Ti^{3+}$ spectral weight in the TEY-XAS spectra in Fig. 1 (*a*), which probe a significant depth into STO, is much smaller than in the core-level photoemission spectra in (*b*), which probe a much narrower vicinity of the interface.

### *Extended-QW scenario*

In view of the 2D-ity of the X-ray generated $V_O$s, the observed 3D-MES in our PA-LAO/STO samples seems puzzling. Which particular kind of the electron states and $V(z)$ shape is needed to realize such 3D-MES? We will start from the well-known picture of the band bending at the ST-LAO/STO interface [4,6,48] sketched in Fig. 3 (*d*). The in-plane $d_{xy}$ orbitals from each $TiO_2$ plane form a sequence of the $d_{xy}$ electron states localized in the out-of-plane direction, whose energies climb $V(z)$ to cross $E_F$ at a finite distance from the interface. The out-of-plane $d_{xz/yz}$ orbitals from each $TiO_2$ plane hybridize into electron states delocalized in the out-of-plane direction, which quantize in $V(z)$ to form a ladder of the the $d_{xz/yz}$ states (where normally only the lowest one is occupied). With no occupied $t_{2g}$ states propagating into the STO bulk, the MES confined in the interfacial $V(z)$ is purely 2D.

One possibility to realize the 3D behaviour (or rather quasi-3D in this particular context) is to increase the depth extension of $V(z)$, which we will call the *extended-QW scenario*. In this case the 3D states will form as the convergence limit of a ladder of 2D states, quantized in the long-range confining potential, as the extension of this potential increases. Such a 2D- to 3D-dimensionality transformation has previously been demonstrated, for example, for multilayer graphene [55] and anatase $TiO_2$ [56]. A similar analysis for the bare STO surface [57] has suggested that whereas a confining potential with an extension of a few monolayers can form a 3D-like ladder of QW states out of the $d_{xy}$ states having large perpendicular $m^*$, it forms only one QW state out of the $d_{xz/yz}$ states with their large interlayer hopping and thus smaller perpendicular $m^*$. This result for the STO surface, however, is at odds with the experimental $k_z$ dispersions demonstrating that, other way around, the $d_{xy}$ states are 2D [18,44] and the $d_{xz/yz}$ ones are 3D [44]. These experimental results go in line with our observations of the 3D-ity of the $d_{xz/yz}$ states in PA-LAO/STO.

Another way to reconcile the extended-QW scenario with our experimental results might be if the MES contribution injected by the top-layer $V_O$s is significantly more delocalized than the intrinsic polar-discontinuity contribution. To check this scenario, we have performed supercell DFT+SIC electronic structure calculations where a 400-atom supercell incorporating 16 $TiO_2$ layers, 6 LaO layers and a (2 × 2) interlayer resolution in a superlattice geometry was employed. As shown in Fig. 5 (*a*), the oxygen deficiency was simulated with a $V_O$ placed in the top $TiO_2$ layer. Fig. 5 (b) presents the results of this calculation (red line) in comparison with the stoichiometric interface (blue). We observe that although the $V_O$ somewhat enhances the asymptotics of the MES towards the STO bulk, it essentially does not change its overall depth extension of ~2 u.c. compared to ST-LAO/STO, where our ARPES experiment has found a purely 2D intrinsic MES.

## *Bulk-metallicity scenario*

Another possibility to realize 3D-MES would be that under X-ray irradiation the whole STO depth within the light absorption depth becomes metallic, and the depth extension of $V(z)$, other way around, shrinks to enable the ARPES experiment directly access the bulk electronic structure. This *bulk-metallicity* scenario is sketched in Fig. 3 (*d*). In our case such short-range $V(z)$ seems more relevant than the long-range one in the previous scenario, because the top-layer $V_O$s release additional MES electrons, as manifested by the observed increase of both $k_F$ and lateral conducting fraction [14] under X-ray irradiation. The increased electron density will then more effectively screen the interfacial-potential discontinuity, reducing the depth over which $V(z)$ saturates to below the probing depth in our ARPES experiment. If upon further progression into the STO bulk the CBM stays below $E_F$, the flat $V(z)$ potential in this region forms the 3D-MES. This scenario explains the 3D-ity of the $d_{xz/yz}$ states observed by the ARPES experiment in PA-LAO/STO. Due to the flat potential, in the STO bulk the $d_{xy}$ states degenerate with the $d_{xz/yz}$. Importantly, the bulk metallicity in STO required for such a scenario can only be realized if the X-ray irradiation causes some effective doping throughout the STO bulk. As we discuss below, such metallicity is essentially the well-known (albeit poorly understood) photoconductivity observed in STO-based systems. The scenario of short-range $V(z)$ combined with the bulk metallicity naturally extends from PA-LAO/STO to bare STO surfaces; the 3D-MES in that case can not be explained only by the $V_O$s [19,20] or any atomic rearrangements [44] in the top STO layer.

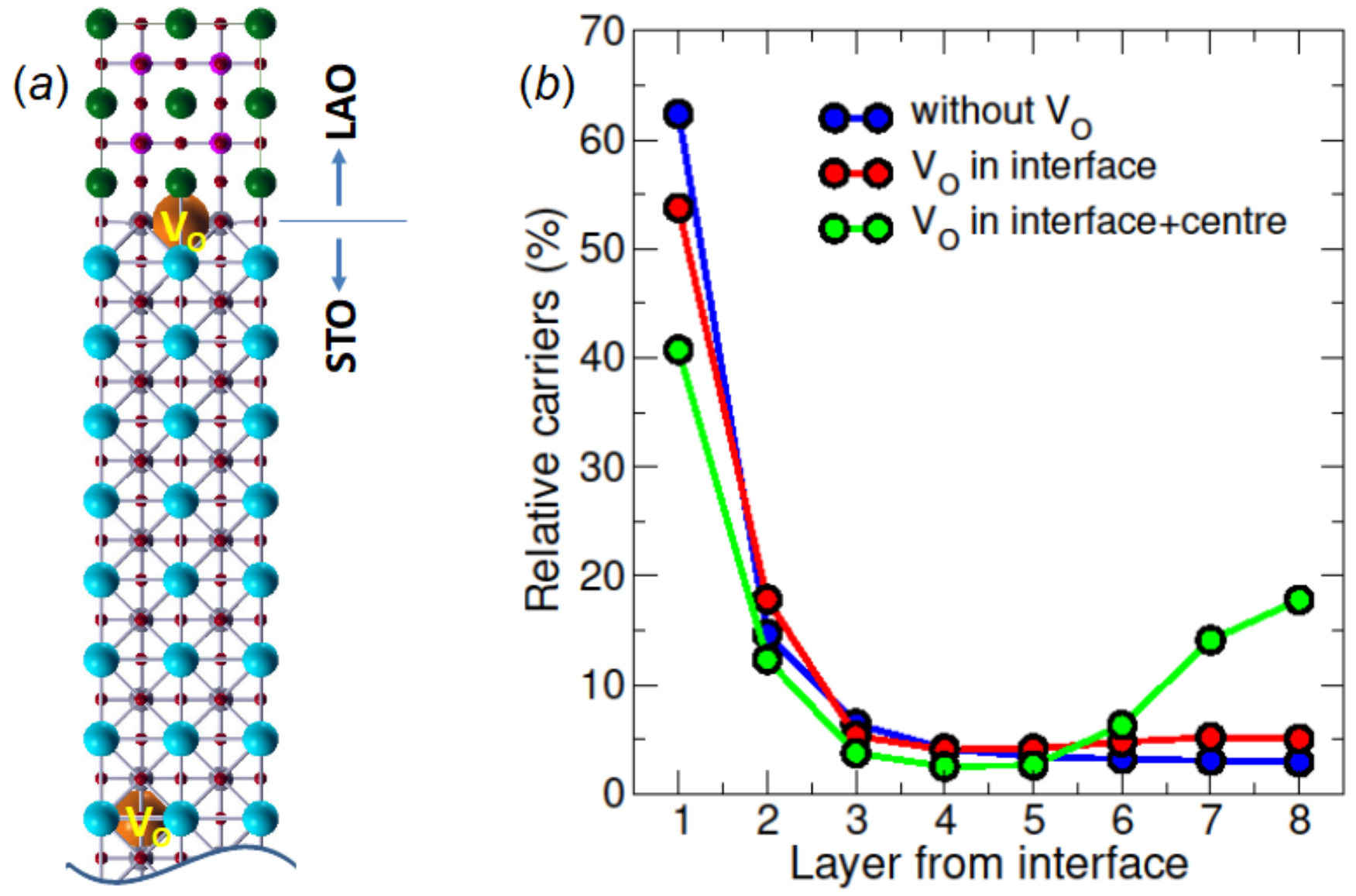

**Fig. 5.** DFT+SIC calculations: (*a*) The supercell incorporating $V_O$s in the top and bulk $TiO_2$ layers; (*b*) Calculated layer-resolved MES density for the stoichiometric and OD-LAO/STO interfaces. Note that only the Ti $t_{2g}$ contribution enters here, since this is the major contribution to the MES.

One possibility for the bulk metallicity of PA-LAO/STO might be that already before the X-ray irradiation the deeper STO layers contained a minute amount of $V_O$s below the sensitivity limit of our ARPES and XAS experiments. Fig. 5 (*b*, green line) shows our calculations which included, in addition to the $V_O$ in the top STO layer, another $V_O$ in the STO-slab centre between the eighth and ninth $TiO_2$ layer, representing the STO bulk. These results indicate that the extension of the MES induced by the bulk $V_O$s is similar to the interfacial ones, restricted within a sphere with a radius of ~2 u.c. The corresponding wavefunctions will start to overlap, forming the 3D character of the MES, at a concentration of only 1.5% which is below the detection limit of our experiment. The previous DFT calculations by Li et al. [10] showed a roughly twice larger spatial extension of the $V_O$-induced MES, with the difference to our calculations plausibly attributed to the neglect of electron correlations. Transport measurements [24] indicate a yet smaller $V_O$-concentration of 0.03% which is necessary to transform STO into 3D metal. Such a minute amount of $V_O$s could likely stay in our samples after their post-annealing. With STO substrates being typically slightly oxygen-deficient, the same mechanism could be at play to form the 3D-MES at bare STO surfaces. We note that the above picture of the photoinduced DX-centers as electron donors forming an extended MES is actually similar to the conventional semiconductors, where tiny dopant concentrations on the promille level can form delocalized electron gas. The exhaustive picture of the OD-LAO/STO interfaces, including the entangled phenomena of the interfacial electrostatic field, $V_O$s, intrinsic and $V_O$-induced MES and photoconductivity, still awaits accurate theoretical description.

The bulk metallicity in STO suggested by this scenario goes in line with the well-known giant photoconductivity of STO-based systems (see, for example, Refs. [32,58][add: Persistent Photoconductivity in 2D electron gases at different oxide interfaces]) that may exceed that in semiconductors by few orders of magnitude [59]. This phenomenon strongly depends on the sample preparation, with the defects including the $V_O$s as well as the cubic-tetragonal phase transition [32] and lattice relaxation [34] playing the main role, and for some LAO/STO [34,60] and OD-STO [59] samples can even become persistent on a time scale of up to days and more. In our experimental setup, we observed the photoconductivity as a drain current to the sample holder through the whole 0.25-mm thickness of our nominally insulating STO substrates which switched on and off under X-ray irradiation. Although the exact physics of photoconductivity in STO yet remains elusive, it is often associated with DX-centers (donor defect) located ~200 meV below the conduction-band minimum. This energy position would allow the photoinduced (in our case X-ray induced) electron states to effectively hybridize with the MES wavefunctions, providing coupling between the MES puddles and thus the 3D-ity of the whole system.

Although the above mechanisms explain the emergence of the 3D electron states, the concentration of $V_O$s and the density-of-states of the corresponding MES in the deeper STO layers stays negligible compared to the $V_O$s and associated states generated by X-rays in the interfacial region. Therefore, these mechanisms can not directly explain the whole large density of the X-ray generated 3D-MES. One possible explanation might be that X-rays generate $V_O$s in STO bulk through creation of Frenkel pairs, where the O atom would become interstitial without the necessity to diffuse over a large distance. This behavior may be facilitated by the repulsion of the X-ray generated mobile electrons from the ones localized at the $V_O$s [34]. Furthermore, the Frenkel-pair mechanism would be consistent with the fact pointed out above that an increase of temperature above ~100K quenches the spectroscopic signatures of the $V_O$s. However, we could not computationally confirm or rule out this scenario within any reasonable computational effort. Full understanding of the X-ray induced formation of the 3D-MES in LAO/STO requires therefore further experimental and theoretical investigations.

## Summary

We used soft-X-ray ARPES to investigate electronic structure of LAO/STO samples grown at low oxygen pressure and post-annealed ex-situ till their (nearly) complete stoichiometric recovery. Under X-ray irradiation at low temperatures below ~100K, the ARPES spectra of these samples show a rapid development of the $Ti^{3+}$ component of the Ti 2*p* core levels and IGS peak, characteristic of the $V_O$s, in parallel with scaling up of the MES spectral weight. Given that at this temperature the oxygen out-diffusion is suppressed, these $V_O$s should be located mostly in the top STO layer. However, the out-of-plane electron dispersions evaluated from the off-resonant ARPES data as well as the variation of the $\mathbf{k}_F$ values over the BZs, evaluated by the gradient method from the Ti 2*p* resonant data, evidence that the X-ray generated MES in our PA-LAO/STO samples is 3D, extending deep into the STO bulk. This behavior, essentially identical to the X-ray generated MES in STO, is clearly distinct from the intrinsic MES in ST-LAO/STO samples, fairly immune to X-ray irradiation, where our ARPES experiment finds pure 2D-ity of the MES. Our DFT+SIC calculations indicate that the spatial extension of the MES induced by $V_O$s in the STO bulk as well as in the top STO layer is about 2 u.c. which can hardly explain the 3D-ty of the MES in PA-LAO/STO. One scenario of this phenomenon, the extended-QW one, involves formation of a dense ladder of QW states in a long-range interfacial potential. More likely in our case is however the bulk-metallicity scenario of the observed 3D-ty, where the X-ray-induced bulk metallicity in STO is accompanied by a short-range interfacial $V$(z) allowing the ARPES experiment to directly access the bulk states. In this case the formation of the 3D-MES may involve the minute concentrations of $V_O$s remnant in the STO bulk and the well-known photoconductivity of STO, and even more exotic mechanisms such as X-ray induced formation of Frenkel pairs. The short-range character of $V(z)$ can be caused by a large concentration of the X-ray generated $V_O$s and thus MES density in the top STO layer, where it effectively screens the interfacial $V(z)$ discontinuity.

## Acknowledgments

We thank C. Cancellieri, R. de Souza and G. Drera for sharing fruitful discussions, and M. Radovic for giving us access to the PLD facility. A.C. acknowledges funding from the Swiss National Science Foundation under grant no. 200021_165529, and F.A. under grant no. 200020B_188709. M.-A.H. acknowledges the support by the Swiss Excellence Scholarship grant ESKAS-no. 2015.0257 and the Romanian UEFISCDI Agency under Contracts No. 475 PN-III-P4-ID-PCE-2020-2540.

## Appendix: Gradient *vs* intensity method of the $k_F$ determination

A crucial element of our further analysis of the ARPES data is accurate determination of the MES' $\mathbf{k}_F$. In fact, its signatures in the ARPES intensity can be strongly distorted not only by sharp variations of the photoemission matrix element but also by broadening of the experimental spectra due incoherent electron scattering (including defects) and limited experimental resolution in $\mathbf{k}$ and $E_b$. Conventionally, one determines from the maxima of the momentum-distribution curve of the spectral intensity at $E_F$ (Fermi intensity $I_F$). However, Straub *et al* [60] have shown that the applicability of this so-called intensity method is restricted to wide-band systems, where the occupied quasiparticle bandwidth $W$ much exceeds $\Delta E$ the experimental resolution, and that in the general case, including narrow-band systems, extremes of the gradient $\mathrm{d}I_W/\mathrm{d}\mathbf{k}$ of the spectral intensity integrated through the whole $W$ resolve $\mathbf{k}_F$ with much higher accuracy. In fact, this so-called gradient method is based on the theoretical many-body definition of the Fermi surface (FS) as restricted by the extremal gradients of the momentum distribution function $n(\mathbf{k})$ [60].

To identify the optimal $\mathbf{k}_F$-determination method in our case, we performed ARPES simulations for a 2D (or 2D cross-section of 3D) parabolic band whose parameters $W$ = 50 meV and $k_F$ = 0.4 Å$^{-1}$ closely resembled the $d_{yz}$ band along $k_x$. With the electron lifetime broadening being negligible close to $E_F$, the corresponding spectral function $A(E,\mathbf{k})$ was modelled as the δ-function centered at the band's dispersion $E(\mathbf{k})$ and terminated at $E_F$. Its ARPES response was simulated by Gaussian convolution with FWHM = 50 meV in the $E$-direction describing the experimental $\Delta E$, and 0.1 Å$^{-1}$ in the $\mathbf{k}$-direction describing the combined effect of disorder and experimental $\mathbf{k}$-resolution.

The results of these simulations, corresponding to the case $W = \Delta E$, are presented in Fig. A1 (b) as the ARPES intensity with the corresponding $I_F(k)$ and $\mathrm{d}I_W/\mathrm{d}k$ curves on top. Whereas the intensity method to determine $k_F$ from the peaks of $I_F$ underestimates its value as much as 14%, the extremes of the gradient d$I_W$/d$k$ yield its value with a remarkable precision of ~0.25%. Important for the gradient-method accuracy is that the spectral intensity is integrated over the whole bandwidth, because the gradient d$I_F$/d$k$ of the Fermi intensity only, also shown in Fig. A1 (b), overestimates the actual $k_F$ by 2.5%. Therefore, our ARPES simulations clearly demonstrate that the gradient method to determine $\mathbf{k}_F$ is optimal in our case.

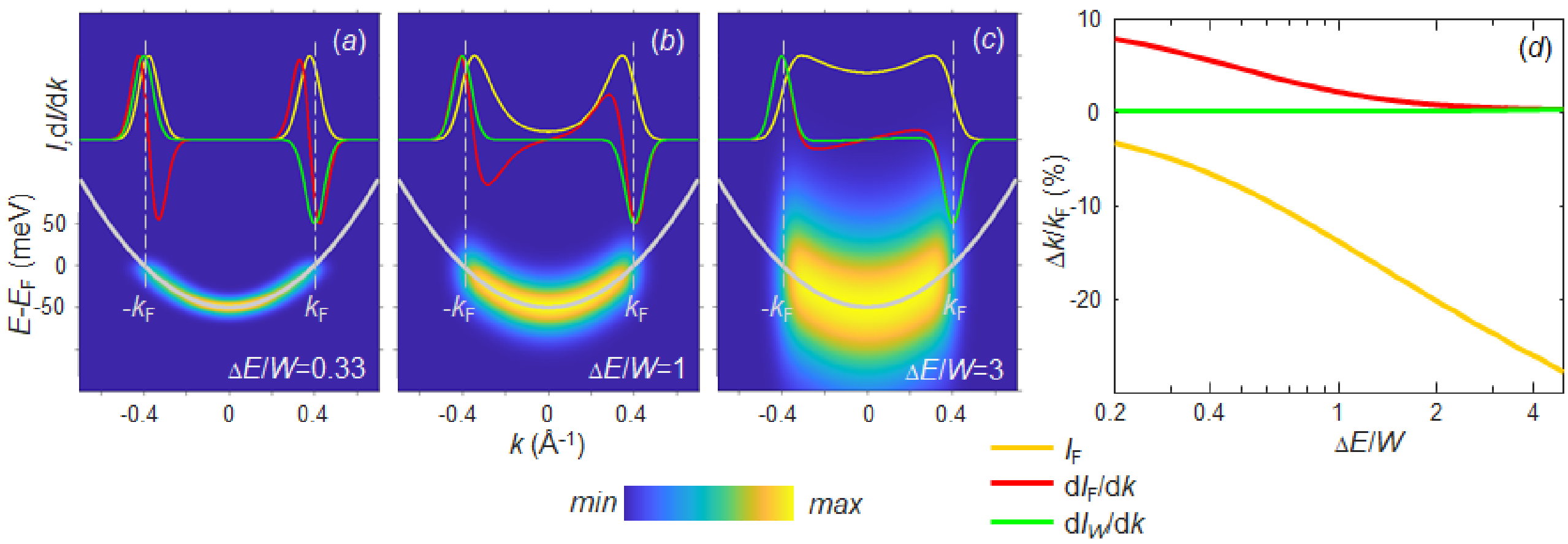

Fig. A1. Determination of $k_F$ at different experimental Δ$E$ vs the occupied bandwidth $W$: ($a$-$c$) Band dispersion superimposed with the ARPES intensity (*bottom*) and the corresponding characteristics of $k_F$ – $I_F(k)$, bandwidth-integrated d$I_W$/d$k$, and d$I_F$/d$k$ – for different Δ$E$/$W$ values; ($d$) Deviation of the peaks in these curves from the true $k_F$ as a function of Δ$E$/$W$. The d$I_W$/d$k$ peaks deliver the most accurate $k_F$ value through all Δ$E$/$W$ values.

Our ARPES simulations extended to smaller and larger Δ$E$/$W$ values are shown in Fig. A1 ($a$) and ($c$), respectively. Obviously, the intensity method improves its accuracy towards small Δ$E$, whereas the $I_F$ gradient does so towards large Δ$E$ compared to $W$. This trend calculated over a wide range of Δ$E$/$W$ is presented in the panel ($d$) which shows the relative deviation of the $I_F(k)$, d$I_W$/d$k$ and d$I_F$/d$k$ extremes from the true $k_F$. A reduction of the spectral **k**-broadening will increase the accuracy of the intensity and $I_F$-gradient methods, in particular of the latter. However, the d$I_W$/d$k$ gradient method stays remarkably accurate through the whole range of Δ$E$ compared to $W$, with a negligible rms deviation of only 0.3% even with the relatively large **k**-broadening used in our simulations. The gradient method to determine $k_F$ can therefore be deemed the most universal. An added advantage of this method is that the narrower width of the d$I_W$/d$k$ peaks compared to $I_F(k)$ makes it less susceptible to the photoemission matrix-element variations with **k**.